\documentclass[prb,showpacs,twocolumn]{revtex4}

\usepackage{amsmath,amsfonts,amssymb,bm}
\usepackage{dcolumn}
\usepackage[final]{graphicx}
\usepackage{bm}

\newcommand*{\E}[0]{\mathcal{E}}
\renewcommand*{\P}[0]{\mathcal{P}}
\newcommand*{\J}[0]{{\bm{\mathcal{J}}}}

\begin{document}

\title{Massive creation of entangled exciton states in semiconductor quantum dots}

\author{Ulrich Hohenester}\email{ulrich.hohenester@uni-graz.at}
\affiliation{Institut f\"ur Theoretische Physik,
  Karl--Franzens--Universit\"at Graz, Universit\"atsplatz 5,
  8010 Graz, Austria}

\date{November 5, 2002}

\begin{abstract}

An intense laser pulse propagating in a medium of inhomogeneously broadened quantum dots massively creates entangled exciton states. After passage of the pulse all single-exciton states remain unpopulated (self-induced transparency) whereas biexciton coherence (exciton entanglement) is generated through two-photon transitions. We propose several experimental techniques for the observation of such unexpected behavior.

\end{abstract}


\maketitle


Entanglement is one of the most intriguing consequences of quantum mechanics which completely lacks a classical counterpart. In particular within the context of the emerging fields of quantum computation~\cite{bouwmeester:00,bennet:00} and quantum communication~\cite{duan:01,gisin:02} it has become clear that entanglement provides the utmost viable element for such future technology, and numerous recent work has been devoted to the preparation and measurement of entangled states in real physical systems. However, the interaction of a quantum system with its environment unavoidably introduces an uncontrollable element to the system's time dynamics, thus spoiling the direct exploitation of entanglement. Noteworthy, such environment losses hitherto seem to be only controllable in a few atomic and photonic systems, but become prohibitively large in the technologically more interesting solid state. Consequently, the identification of long-lived and sufficiently well protected solid-state excitations has now turned into the most pertinent issue within this area of research.

In this respect, semiconductor quantum dots,~\cite{woggon:97,hawrylak:98,bimberg:98} or {\em artificial atoms}\/ as they are sometimes called because of their atomic-like carrier states, provide a promising new class of material, which resembles many of the atomic properties whilst offering at the same time all the flexibility of semiconductor nanostructures. Quantum dots consist of a small island of lower-bandgap material embedded in a solid-state matrix of higher-bandgap material. Proper choice of the material and dot parameters thus allows the confinement of a few carrier states within this lower-bandgap region, resulting in discrete spectra and strongly enhanced lifetimes: indeed, remarkably long dephasing times have been recently reported for optical excitations (excitons) which were solely governed by radiative decay.~\cite{borri:01} Another property that has attracted enormous interest is the possibility to create several electron-hole pairs (multi-excitons) within a single quantum dot, where, because of Coulomb renormalziations, the spectra exhibit a surprisingly rich fine structure.~\cite{zrenner:00,bayer:00a} We emphasize that it is precisely this Coulomb correlation effect which is at the heart of quantum-dot based single-photon sources~\cite{gerard:99,michler:00} and which recently allowed for an optically induced exciton entanglement.~\cite{chen:00,chen:02}

In this paper it is shown that a strong laser pulse propagating in a macroscopic sample of inhomogeneously broadened quantum dots massively creates entangled exciton states. Such transition is due to the above-mentioned peculiarities of quantum dots (long lifetimes and Coulomb renormalizations) and is mediated by the coherent light-matter interaction within a self-modulation process, thus occurring under very general conditions. Quite generally, two basic phenomena are made responsible: firstly, above a given power threshold a laser pulse can propagate in a system of inhomogeneously broadened two-level systems without suffering significant losses (self-induced transparency);~\cite{mccall:67,mccall:69} secondly, in case of two-photon resonance population can be directly channeled between the ground- and biexciton state without populating the intermediate exciton states.~\cite{brunner:94}


Our theoretical approach is based on the simulation of the coupled light-matter system, which requires the solution of both the material and Maxwell equations: here, the laser pulse (described through its electric field $\bm\E$) creates an interband polarization in the quantum dots, which, on its part, serves again as a source term in Maxwell's equation and thus acts back on $\bm\E$. Let us first discuss the time dynamics of a single dot (introducing an appropriate ensemble average later) which we describe within a common master-equation framework.~\cite{walls:95,scully:97} Following Refs.~\onlinecite{panzarini.prb:02,hohenester:02} we characterize the quantum dot system through its density-matrix $\bm\rho$, whose diagonal elements $\bm\rho_{xx}$ describe the occupation of the few-particle states $x$ (groundstate, single- and multi-excitons), and the off-diagonal terms $\bm\rho_{xx'}$ account for the coherence between states $x$ and $x'$. The time dynamics of $\bm\rho$ is then governed by~\cite{panzarini.prb:02,hohenester:02} ($\hbar=1$ throughout):

\begin{equation}\label{eq:master}
  \dot{\bm\rho}=-i(\bm h_{\rm eff}\bm\rho-\bm\rho\bm h_{\rm eff}^\dagger)
  +\J\bm\rho,
\end{equation}

\noindent with $\bm h_{\rm eff}=\bm h_o+\bm h_{op}-i\bm\Gamma$ accounting for: $\bm h_o$, the Coulomb-renormalized few-particle states $x$; $\bm h_{op}$, the light-coupling described within the usual rotating-wave and dipole approximations;~\cite{scully:97} $i\bm\Gamma$, dephasing and relaxation due to environment interactions; finally, $\J$ accounts for in-scatterings which guarantee that the trace of $\bm\rho$ is preserved at all times.\cite{panzarini.prb:02,hohenester:02} In this paper we shall consider low temperatures throughout, and thus take spontaneous photon emissions as the only source of dephasing and relaxation.~\cite{borri:01,remark:dephasing}

As regarding the time evolution of the light pulse, we assume the geometry depicted in Fig. 1a of a laser pulse entering from the left-hand side into the sample of inhomogeneously broadened quantum dots. Denoting the pulse propagation direction $z$ and assuming an electric-field profile $\bm\E_o\cos\omega_o t$, with envelope $\bm\E_o$ and central frequency $\omega_o$, we describe the light propagation in the slowly-varying envelope approximation:~\cite{panzarini.prb:02,mandel:95}

\begin{equation}\label{eq:maxwell}
  \left(\partial_z+\frac n c\partial_t\right)\bm\E_o(z,t)\cong
  -\frac{2\pi\omega_o}{nc}\;\mbox{Im}\bm\P(z,t),
\end{equation}

\noindent where $n$ is the semiconductor refraction index and $c$ the speed of light. Most importantly, the term on the right-hand side describes the back-action of the material polarization $\bm\P(z,t)$ on the light propagation. Here:~\cite{panzarini.prb:02,hohenester:02}

\begin{equation}\label{eq:ensemble}
  \bm\P(z,t)=\mathcal{N}\int g(\epsilon)d\epsilon\;\sum_{xx'} 
  \bm M_{x'x}(\epsilon)\bm\rho_{xx'}(\epsilon,z;t),
\end{equation}

\noindent with ${\cal N}$ the uniform dot density, $\epsilon$ the exciton energies, $g(\epsilon)$ a normalized distribution characterized through the full-width of half maximum $\delta^*$ of the inhomogeneously broadened ensemble, and $\bm M_{x'x}$ the optical dipole matrix elements. Note that for each $z$ and $\epsilon$ the time evolution of $\bm\rho_{xx'}(\epsilon,z;t)$ is given by Eq.~(\ref{eq:master}).

\begin{figure}
\includegraphics[width=0.85\columnwidth]{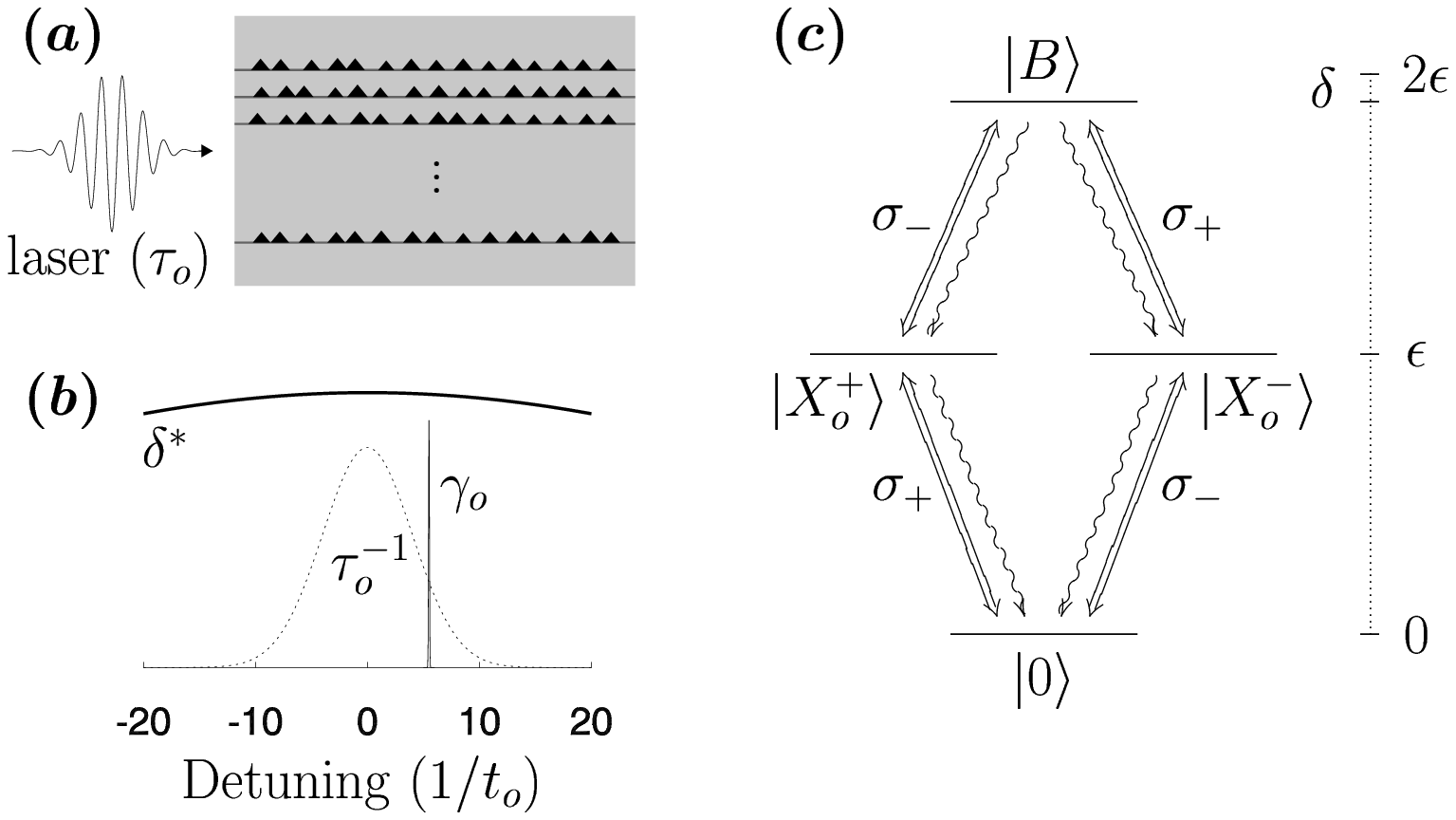}
\caption{
Schematic representation of: (a) the setup, where an intense laser pulse with a temporal width $\tau_o$ enters from the left-hand side into the dot sample; (b) the groundstate exciton absorption spectrum for the inhomogeneously broadened dots ($\delta^*$) and for a single dot ($\gamma_o$); the dashed line shows the spectral width of the laser pulse; (c) generic quantum-dot level scheme (for discussion see text); optical selection rules for circularly polarized light ($\sigma_\pm$) and spontaneous photon emission processes (wiggled lines) apply as indicated.~\cite{remark:level-scheme}
}
\end{figure}

\begin{figure*}
\includegraphics[height=2.8in]{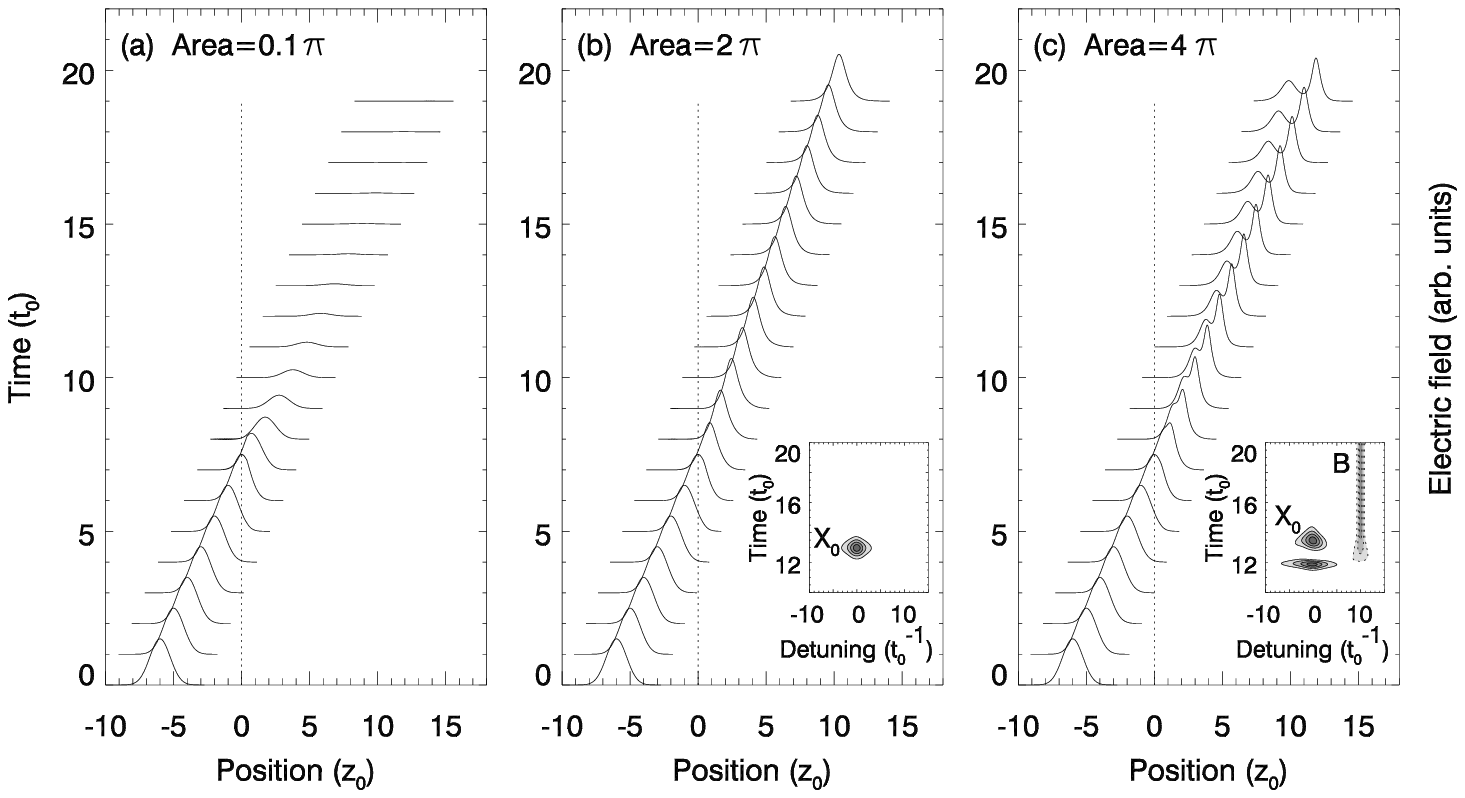}
\caption{
Results of our simulations of pulse propagation in a sample of inhomogeneously broadened quantum dots and for different pulse areas; we use linear polarization and assume a setup where the pulse enters from a dot-free region (negative $z$-values) into the dot region. The insets show contour plots of the time evolution of $\bm\rho(\epsilon,z;t)$ for $z=5 z_o$ (with detuning $\epsilon-\omega_o$). We use a prototypical biexciton binding of 20 $t_o^{-1}$.
}
\end{figure*}


In our calculations we assume a laser frequency $\omega_o$ tuned to the maximum of the inhomogeneously broadened exciton-groundstate transitions, Fig. 1b, and a typical exciton energy splitting of the order of several tens of meV,~\cite{hawrylak:98,bimberg:98} thus restricting our analysis to the generic level scheme of Fig.~1c, which consists of: the groundstate $|0(\epsilon)\rangle$ (no electron-hole pairs present); the spin-degenerate excitons of lowest energy, $|X_o^\pm(\epsilon)\rangle$; and the biexciton groundstate $|B(\epsilon)\rangle$, whose energy $2\epsilon-\delta$ is reduced because of Coulomb renormalizations.~\cite{zrenner:00,bayer:00a,hohenester:02} For typical values of $\delta^*\sim 20$ meV and $\gamma_o\sim 1$ $\mu$eV (Ref.~\onlinecite{borri:01}) for the inhomogeneous and homogeneous (lifetime) broadenings, respectively, and assuming laser pulses with $\tau_o\sim 1$--10 ps, one immediately observes that:

\begin{equation}\label{eq:sit}
  \delta^*\gg \tau_o^{-1}\gg \gamma_o.
\end{equation}

\noindent In Ref.~\onlinecite{panzarini.prb:02} we made the important observation that inequalities (\ref{eq:sit}) have severe consequences for the pulse propagation. Figure 2 shows results of our simulations based on Eqs.~(\ref{eq:master}--\ref{eq:ensemble}) for different pulse areas $\int_{-\infty}^\infty dt\;\mu_o\E_o(t)$, with $\mu_o$ the dipole moment of the bulk semiconductor. Consider first the case of a weak laser pulse entering the dot region, Fig. 2a, whereby the laser excites excitons and suffers attenuation; a more detailed analysis reveals exponential damping (Beer's law of linear absorption) with $z_o=nc/ (2\pi^2\mathcal{N}\omega_o\mu_o^2g(\omega_o))$ providing a characteristic length scale~\cite{mandel:95,mccall:67,mccall:69} (henceforth we shall measure length in units of $z_o$, time in units of $t_o=n z_o/c$, and energy in units of $t_o^{-1}$, with  $z_o\sim 250$ $\mu$m, $t_o\sim 3$ ps, and $t_o^{-1}\sim 0.2$ meV for typical InGaAs dot samples~\cite{panzarini.prb:02,borri:02a}). Because of the weak dephasing the laser-induced coherence keeps stored in the material even after attenuation of the laser pulse. As consequence, when the laser intensity is further increased, Fig. 2b, this stored energy can again be fully extracted from the material and given back to the laser pulse, as first demonstrated in the seminal work of McCall and Hahn.~\cite{mccall:67,mccall:69} This leads to a propagation where at each instant of time the pulse gives and receives the same amount of energy from the material  (self-induced transparency). While, strictly speaking, such ideal performance is only expected for a generic two-level scheme,~\cite{mccall:67,mccall:69} the results of Fig.~2 clearly demonstrate that all essential features, such as stable pulse propagation or pulse breakup at the highest field strengths, remain in case of the more complicated level scheme of Fig.~1c.

However, in the inset of Fig. 2c and even more clearly in Fig. 3 we observe that after passage of the pulse some biexciton population remains. The underlying states are {\em entangled exciton states}.\/ To see that, we first note that because of the negligible dephasing the time dynamics can be considered as almost coherent (thus allowing for a wavefunction description); introducing furthermore the suggestive notations $|00\rangle$ for the groundstate, $|10\rangle$ and $|01\rangle$ for the excitons $|X_o^\pm\rangle$, and $|11\rangle$ for the biexciton,~\cite{remark:level-scheme} whereby we have assumed that in the strong-confinement regime $|B\rangle$ is approximately given by the product state $|X_o^+\rangle \otimes |X_o^-\rangle$,~\cite{bayer:00a} within each of these dots the wavefunction is of the form:~\cite{remark:wavefunction}

\begin{equation}\label{eq:entangled}
  |\Psi\rangle=const\times\left(|00\rangle+\xi|11\rangle\right),
\end{equation}

\noindent with $\xi$ a complex number (see Fig.~3)---this wavefunction is exactly an {\em entangled}\/ one~\cite{bouwmeester:00,remark:entanglement} (we checked that our results do not depend decisively on the specific values of $\tau_o$ and $\delta$, and thus reflect a general behavior). To understand the origin of this entanglement-creation, we note that the transition occurs at the photon energy where the biexciton is in two-photon resonance, i.e., $2\epsilon-\delta=2\omega_o$. Assuming that for linear polarization the level scheme of Fig.~1c reduces to an effective three-level one (since only one of the superpositions $(|X_o^+\rangle\pm|X_o^-\rangle)/\sqrt 2$ couples to the light), at two-photon resonance the effective Hamiltonian of Eq.~(\ref{eq:master}) reduces to:~\cite{remark:level-scheme}

\begin{equation}\label{eq:heff}
  \bm h_o+\bm h_{op}=-\frac 1 2 \left(
  \begin{array}{ccc}
    0 & \Omega & 0 \\
    \Omega & -\delta & \Omega\\
    0 & \Omega & 0 \\
  \end{array}\right),
\end{equation}

\begin{figure}
\includegraphics[width=0.75\columnwidth]{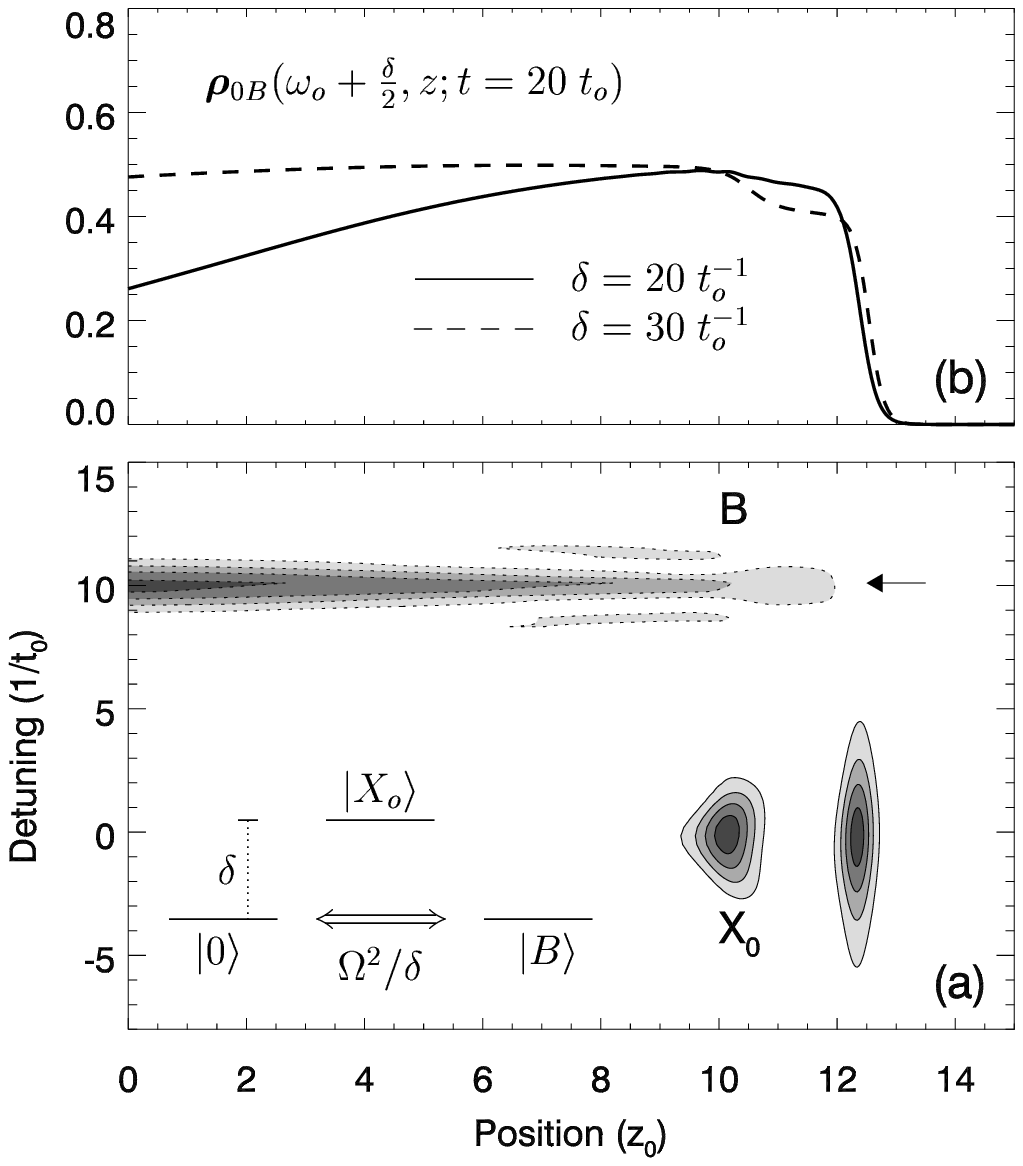}
\caption{
(a) exciton and biexciton population at $t=20 t_o$ (contour lines of 0.25, 0.45, 0.65, and 0.85, respectively); (b) groundstate-biexciton coherence $\bm\rho_{0B}(\omega_o+\frac\delta 2,z;t=20\;t_o)$ at two-photon resonance (see arrow)---note that a value of $\frac 1 2$ corresponds to a maximally entangled Bell state;~\cite{remark:entanglement} the value of $\bm\rho_{0B}$ can be controlled through variation of $\delta$ (dashed line) or, equivalently, of $\tau_o$.}
\end{figure}

\noindent whereby we have used a rotating-frame representation according to $\omega_o|X_o\rangle\langle X_o|+2\omega_o|B\rangle\langle B|$,~\cite{panzarini.prb:02} and $\Omega=\mu_o\E_o$ is the usual Rabi frequency. Eq.~(\ref{eq:heff}) describes a system where the two states $|0\rangle$ and $|B\rangle$ are coupled by $\Omega$ through an auxiliary and off-resonant level $|X_o\rangle$ (see inset of Fig.~3a). For constant $\Omega$ and assuming $\Omega\ll \delta$ one can analytically obtain the eigenstates of Eq.~(\ref{eq:heff}),~\cite{barnett:97} which consist of: the bare exciton state $|X_o\rangle$; two mixed states of $|0\rangle$ and $|B\rangle$. More specifically, if the system is initially in state $|0\rangle$ its time evolution is governed by the Hamiltonian:~\cite{barnett:97}

\begin{equation}\label{eq:2p}
  \bm h_o+\bm h_{op}\cong \frac {\Omega^2}{\delta} \left( 
    |B\rangle\langle 0|+
    |0\rangle\langle B|\right).
\end{equation}

\noindent Apparently, as time goes on the system will oscillate between $|0\rangle$ and $|B\rangle$; consequently, the final biexciton population in Figs.~2 and 3 is governed by the pulse intensity and can be controlled through variation of $\tau_o$.

Although the basic mechanisms underlying the creation of such exciton entanglement, i.e., excitation and de-excitation of single excitons (Rabi-type oscillations) and two-photon transitions, are essentially single-dot effects, the entanglement creation considered here is a genuine cooperative phenomenon. First note that while populating biexcitons the laser pulse looses intensity: if we would neglect the possibility of pulse reshaping such losses would be accompanied by a reduction of the pulse area, which would result in a rapid pulse attenuation (since the condition for self-induced transparency would no longer be fulfilled); however, because of the back-action of the material polarization on $\bm\E$, Eq.~(\ref{eq:maxwell}), at each instant of time the pulse reshapes to conserve area and to compensate for the losses suffered (see Ref.~\onlinecite{mccall:69} for a related discussion). Thus, self-induced transparency at the single-exciton level and biexciton creation are not independent phenomena, but just represent two different facets of a single cooperative phenomenon.

We envision a number of experimental techniques for the measurement of such massive entanglement. Firstly, we propose to monitor the luminescence after passage of the laser pulse: since in absence of the strong laser fields the excited biexciton states must relax via the interconnecting exciton states, Fig.~1c, the luminescence spectra consist of two peaks centered around $\omega_o\pm \delta/2$ (and no signal at $\omega_o$). Secondly, consider a coherent-carrier control setup of a weak probe pulse with frequency $\omega_o-\delta/2$ following the first pulse: for an appropriately chosen phase difference between the pulses, the probe pulse can propagate over long distances without suffering losses, whereby the attenuation due to exciton creation is compensated by the gain through stimulated biexciton emission. Finally, we propose to measure the photon noise: let us take in Eq.~(\ref{eq:2p}) the opposite viewpoint and consider the material polarization as a semiclassical source and the light field as the quantum system. Then,~\cite{walls:95,scully:97}

\begin{equation}\label{eq:squeezing}
  \bm h_\ell\cong 
  \frac{1}{2\pi\delta} \left(\lambda\;   a_+^\dagger a_-^\dagger+
                             \lambda^*\; a_+a_-\right)
\end{equation}

\noindent is the Hamiltonian for the light field, with $\lambda\cong\int d\epsilon\; dz\;\bm\rho_{B0}(\epsilon,z)$ and $a_\sigma^\dagger$ the creation operator for a photon with polarization $\sigma$. Most importantly, a Hamiltonian of this form is known to lead to multi-mode squeezing~\cite{walls:95,scully:97,caves:85} (for the parameters considered and assuming $z\sim10 z_o$ we estimate a squeezing factor of 1--10 percent~\cite{caves:85}). Thus, while creating entangled exciton states the light field of the laser pulse becomes squeezed, which should be observable in heterodyne detection~\cite{haus:00} as a direct signature of the entangled states. Future work will also address pulse propagation in waveguide structures and cavities where further control of the biexciton population is possible, which might be of relevance for possible quantum-communication applications (e.g., multiparticle entanglement) or non-classical light sources.


I am indebted to Giovanna Panzarini who substantially contributed at an early stage of this work. Her memory will keep alive in our hearts. Elisa Molinari is acknowledged for continuous support and helpful discussions.


\begin{thebibliography}{30}
\bibitem[{rem({\natexlab{b}})}]{remark:level-scheme}
\bibinfo{note}{Note that our approach remains valid for linearly polarized
  exciton states, provided that both states couple with comparable strength to
  the light and for exciton splittings much smaller than the biexciton binding
  $\delta$.}
\bibitem[{rem({\natexlab{c}})}]{remark:wavefunction}
\bibinfo{note}{The total wavefunction is thus given by the product
  $\prod_\ell\left(|00\rangle_\ell+\xi_\ell|11\rangle_\ell\right)$, where
  $\ell$ runs over all dots whose biexciton energy is in two-photon resonance,
  i.e., $2\epsilon-\delta=2\omega_o$.}
\bibitem[{rem({\natexlab{d}})}]{remark:entanglement}
\bibinfo{note}{We emphasize that the results of Fig.~3 unambiguously prove the
  exciton entanglement. First note that the constant prefactor in
  Eq.~(\ref{eq:entangled}) is $(1+\xi^2)^{-\frac 1 2}$, where for simplicity we
  have assumed $\xi$ to be real. Next, with the pure state (\ref{eq:entangled})
  the off-diagonal components of the density matrix $\bm\rho_{0B}$ become
  $\langle 0|\Psi\rangle\langle\Psi|B\rangle=\xi/(1+\xi^2)$. Thus, a value of
  $\frac 1 2$ corresponds to a maximally entangled Bell state, whereas smaller
  values indicate a reduced degree of entanglement. Finally note that within
  the slowly-varying envelope approximation at resonance $\bm\rho_{0B}$ is a
  real quantity (as shown in Fig.~3b).}
\end{thebibliography}
\end{document}